\pdfoutput=1
\documentclass[aps,twocolumn, floatfix, prb]{revtex4}
\usepackage{graphicx}
\DeclareGraphicsExtensions{.pdf}
\usepackage{amsmath,amssymb,bbold,bm}
\usepackage{float}
\usepackage{epstopdf}

\newcommand{\br}{{\bm r}}

\newcommand{\bS}{{\bm S}}

\newcommand{\cH}{{\cal H}}

\newcommand{\cE}{{\cal E}}

\def\sgn{\mathop{\rm sgn}\nolimits}

\begin{document}

\title{Self-organized topological state in the
magnetic chain on the surface  of a superconductor}

\author{I. Reis}
\author{D.J.J. Marchand}
\author{M. Franz}
\affiliation{Department of Physics and Astronomy, University of
British Columbia, Vancouver, BC, Canada V6T 1Z1}
\affiliation{Quantum Matter Institute, University of British Columbia, Vancouver BC, Canada V6T 1Z4}

\begin{abstract}
Electronic states associated with a chain of magnetic adatoms on the surface of an ordinary $s$-wave superconductor have been shown theoretically to form a one dimensional topological phase with unpaired Majorana fermions bound to its ends. In a simple 1D effective model the system exhibits an interesting self-organization property: the pitch of the spiral formed by the adatom magnetic moments tends to adjust itself so that electronically the chain remains in the topological phase whenever such a state is physically accessible. Here we examine the physics underlying this self-organization property in the framework of a more realistic 2D model of a superconducting surface coupled to a 1D chain of magnetic adatoms. Treating both the superconducting order and the magnetic moments selfconsistently we find that the system retains its self-organization property, even if the topological phase extends over a somewhat smaller portion of the phase diagram compared to the 1D model. We also study the effect of imperfections and find that, once established, the topological phase survives moderate levels of disorder.  

\end{abstract}

\date{\today}
\maketitle

\section{Introduction}
Even though tantalizing experimental signatures of Majorana particles have been reported in semiconductor quantum wires \cite{mourik1,das1,deng1,rohkinson1,finck1,churchill1,Ramon}, there has been much less tangible progress to date in exploring and confirming their exotic physical properties or harnessing their potential for technological applications. Further progress will be achieved by perfecting the existing devices and by engineering new systems which harbor unpaired Majorana zero modes and are amenable to a wide range of experimental probes. Indeed there is no shortage of theoretical proposals to implement and probe Majorana particles in solid state systems \cite{alicea_rev,beenakker_rev,stanescu_rev,elliot_rev}.

\begin{figure}[b]
\includegraphics[width=7cm]{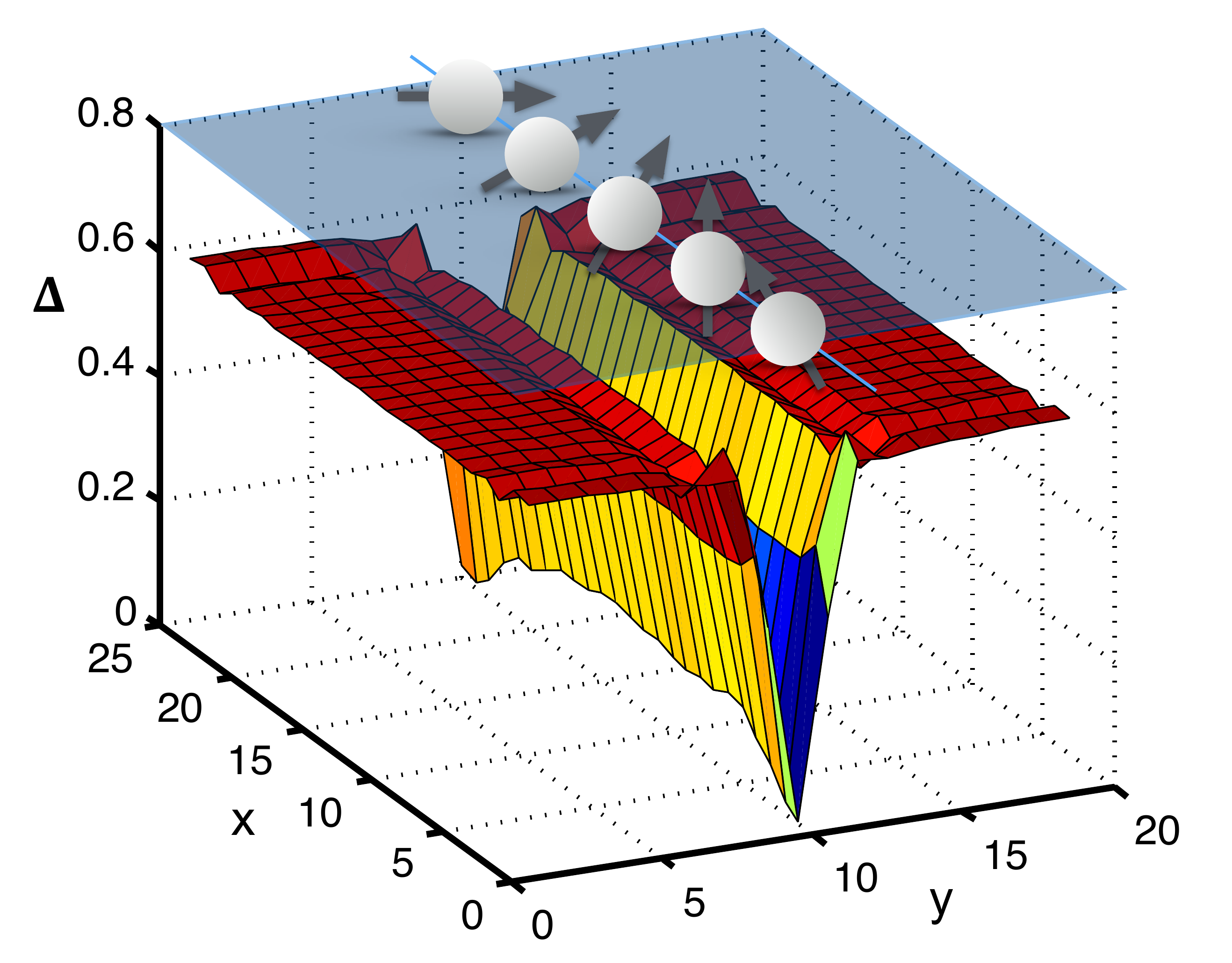}
\caption{Schematic depiction of the system: adatoms with magnetic moments forming a spiral on the surface of a superconductor.  Suppression of the superconducting order parameter $\Delta(\br)$, calculated as described in Section III below, is also shown. } \label{fig1}
\end{figure}

A specific system that we study in this paper consists of a chain of magnetic atoms, such as Fe, Cr or Gd, deposited on the atomically flat surface of an ordinary $s$-wave superconductor such as Pb or Nb, as illustrated in Fig.\ \ref{fig1}.  It has been shown previously that if the magnetic moments in the adatom chain form a spiral with the correct pitch, the Shiba states \cite{lu1,shiba1,rusinov1,balatsky_rev} (that form inside the SC gap in response to the magnetic moments) can give rise to an effective 1D topological superconductor (TSC) with Majorana fermions bound to its ends \cite{choy1,martin1,nadj-perge2,pientka1,pientka2}. In practice such systems can be engineered and probed by scanning tunneling microscopy \cite{nadj-perge1} and preliminary spectroscopic evidence for zero modes has indeed been reported\cite{yazdani00}. Unlike other proposals for Majorana fermions this system does not rely on a strong spin-orbit coupling (SOC) to produce a 1D topological state. Instead, the exchange coupling of the electrons to the magnetic moments mimics the effect of SOC when viewed in the reference frame rotating along with the spiral. Thus, the parameter relevant to the topological phase is not the SOC strength $\lambda$ but the exchange coupling $J$ which can be large in solids, potentially leading to a more robust protection of the Majorana zero modes. Recently, a convenient way to effect Majorana fermion braiding in this setup has been proposed \cite{neupert1}.

Another interesting feature of the system is its self-organization property
\cite{simon1,loss1,vazifeh1}. As we shall review in more detail below, in a  model describing the magnetic moments interacting with a superconductor,  the spiral pitch $G$ must be in a certain relationship with the electron chemical potential $\mu$ for the system to exhibit the topological phase. The spiral pitch $G$ in turn depends on the coupling between the moments that is typically mediated by the electrons in the substrate. A priori it is not at all clear that the resulting equilibrium $G$ will have the correct value to bring about the topological phase. It is found \cite{simon1,loss1,vazifeh1}, however, that at least in a simple 1D model the spiral pitch that minimizes the system energy is exactly the one required for the topological phase to emerge  -- remarkably, the system {\em wants} to be topological.

Recently, it has been argued \cite{bela1} that the self-organization property described above might only be effective in a purely 1D model and may thus not be relevant to a real experimental system that involves a 2D superconducting surface. The key insight behind Refs.\ \onlinecite{simon1,loss1,vazifeh1} is that the spiral ordering vector $G$ will coincide with $2k_F$ of the underlying normal electron gas because in 1D its static spin susceptibility $\chi(q)$ peaks at $q=\pm 2 k_F$. This behavior is only slightly modified by the inclusion of the superconducting order. In a 2D electron gas, however, $\chi(q)$ has only an inflection point at $2k_F$ and there is thus no reason to expect spiral ordering at $G\approx 2 k_F$. The authors of Ref.\ \onlinecite{bela1} suggested introducing SOC into the 2D problem to restore the spiral ordering. They also pointed out that small amounts of disorder, especially in the adatom positions, could destroy the spiral ordering and thus the topological phase.

With the goal of addressing the fate of the self-organization property away from a simple 1D model we perform a study here of a realistic lattice model (similar to Ref.\ \onlinecite{nadj-perge2}) describing a 2D superconducting surface with electrons coupled to a 1D chain of magnetic moments positioned commensurately with the ionic lattice of the substrate. We consider both periodic and open ended chains, solving selfconsistently for the SC order parameter and the spiral pitch $G$. Our main finding is that the self-organization property previously established in a purely 1D model \cite{simon1,loss1,vazifeh1} persists more or less intact in a model that takes into account the 2D geometry of the SC surface. A simple intuitive picture behind this result can be given as follows. As is well known from the physics of 1D quantum wires \cite{sau2,oreg1}, for the system to be in the topological phase the magnetic gap must dominate over the SC gap $\Delta$. In our present setup this translates into the requirement that the exchange coupling $J$ must be larger than the SC pairing scale $\Delta$. In this limit, each magnetic moment acts as a strong pair breaking defect. A line of such strong pair breakers then locally suppresses $\Delta(\br)$ essentially to zero as illustrated in Fig. \ref{fig1}. This ``trench'' in the SC pair potential hosts low-energy electron states that can be viewed as a 1D electron system  described effectively by the same model as discussed in  Refs.\ \onlinecite{simon1,loss1,vazifeh1}. The self-organization property of the 1D system is therefore recovered at the energy scales below the bulk superconducting gap $\Delta_0$.

We also study the effects of disorder and find that it has only a mild effect on the stability of the topological phase. In a realistic system magnetic adatoms will be registered to the minima of the local surface potential which is necessarily commensurate with the underlying ionic lattice of the substrate. For this reason we feel that the positional disorder of the type emphasized in Ref.\ \onlinecite{bela1} should not be relevant to the physical system. In our lattice model we study the substrate disorder (modeled by local variations in the chemical potential $\mu$) and disorder in the exchange couplings $J$. Weak and moderate strengths of these lead to local deviations away from the perfect spiral order which however do not affect the Majorana end states. Strong disorder causes a proliferation of domain walls in the spiral order and the eventual destruction of the topological phase.

\section{The Model and its basic properties}

\begin{figure}[b]
\includegraphics[width=8.7cm]{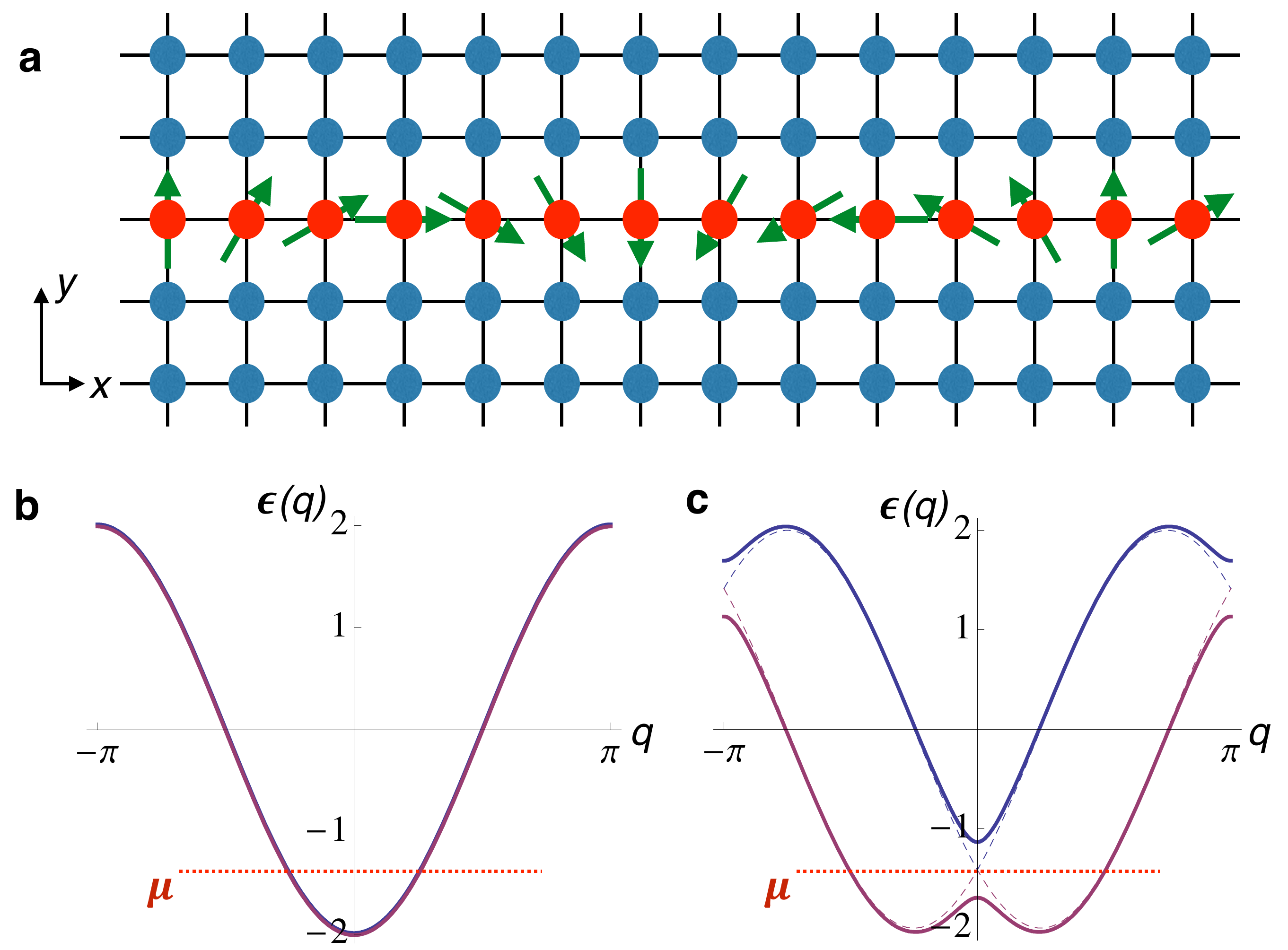}
\caption{a) The tight binding lattice with a row of magnetic moments used in our numerical simulations. b) The normal state electron spectrum in the 1D limit for $J=0$ and c) for $J>0$.  } \label{fig2}
\end{figure}

We describe the surface of a superconductor  by a tight-binding model for electrons on the square  lattice coupled to magnetic moments $\bS_i$ associated with adatoms. The normal state is described by the Hamiltonian
\begin{equation}\label{hspi1}
{\cal H}_0=-\sum_{ij\sigma}t_{ij} c^\dagger_{i\sigma} c_{j\sigma} - \mu\sum_{i\sigma}  c^\dagger_{i\sigma} c_{i\sigma}
 +J\sum_{i\in I}  \bS_i\cdot(c^\dagger_{i\sigma}{\bm \sigma}_{\sigma\sigma'} c_{i\sigma'})
\end{equation}
Here $c^\dagger_{j\sigma}$ creates an electron with spin $\sigma$ on site $j$ of the square lattice containing $N=L_x\times L_y$ sites,
$J$ stands for the exchange coupling constant and ${\bm \sigma}=(\sigma^x,\sigma^y,\sigma^z)$ is the vector of Pauli spin matrices. Magnetic moments of adatoms $\bS_i$ are assumed to live on a single row of lattice sites denoted by $I$ as indicated in Fig.\ \ref{fig2}a. We take the surface to lie in the $xy$ plane with the line of adatoms along the $x$ direction. In the following we shall consider periodic boundary conditions along $y$ and both periodic and open boundary conditions along $x$. We also note that for $L_y=1$ one recovers the 1D model studied in Refs.\ \onlinecite{simon1,loss1,vazifeh1}.

Superconducting order is introduced by assuming an on-site attractive interaction between electrons parametrized by $V>0$
\begin{equation}\label{hspi1a}
{\cal H}={\cal H}_0-V \sum_j n_{j\uparrow}n_{j\downarrow},
\end{equation}
with $n_{j\sigma}=c^\dagger_{j\sigma}c_{j\sigma}$ the number operator. We treat the interaction within the  standard Bogoliubov-de Gennes (BdG) formalism by decoupling the four-fermion term in the particle-particle channel. This leads to the second quantized BdG Hamiltonian of the form
\begin{equation}\label{hspi1b}
{\cal H}={\cal H}_0+ \sum_j\left (\Delta_j c^\dagger_{j\uparrow} c^\dagger_{j\downarrow}+{\rm h.c.}-{1\over V}|\Delta_j|^2\right).
\end{equation}
with 
\begin{equation}\label{hspi1c}
\Delta_j=V\langle c_{j\uparrow} c_{j\downarrow} \rangle
\end{equation}
 the SC order parameter. The expectation value in Eq.\ (\ref{hspi1c}) is taken in the ground state of the BdG Hamiltonian (\ref{hspi1b}). In the following Section we will solve Eqs.\ (\ref{hspi1b}) and (\ref{hspi1c}) by selfconsistent iteration for various system geometries, boundary conditions and magnetic moment configurations.

In Refs.\ \onlinecite{simon1,loss1,vazifeh1} it was assumed that the adatom  magnetic moments form a coplanar spiral at wavevector $G$, 
\begin{equation}\label{hspi2}
\bS_j=S[\cos{(Gx_j)},-\sin{(Gx_j)},0],
\end{equation}
where $(x_j,y_j)$ are the coordinates of site $j$ . We note that since the model under consideration has a full SU(2) spin symmetry, assuming the moments to rotate in the $xy$ plane does not lead to any loss of generality. By a simultaneous global SU(2) rotation of the magnetic moments and the electron spins one can rotate the spiral into an arbitrary plane. For the most part in the following we shall continue using Eq.\ (\ref{hspi2}) with $G$ a free parameter to describe the magnetic moments. In Section IIIB, where we study the effect of disorder,  we will however relax this assumption and let each moment $\bS_j$ equilibrate individually. For a clean system this procedure confirms in an unbiased way that a coplanar spiral state indicated in Eq.\ (\ref{hspi2}) corresponds to the true ground state of the system.

For a spiral configuration of the adatom magnetic moments given by Eq.\ (\ref{hspi2}) and with periodic boundary conditions along $x$ it is expedient to perform a spin-dependent gauge transformation  \cite{martin1},
\begin{equation}\label{hspi3}
c_{j\uparrow}\to c_{j\uparrow} e^{{i\over 2}Gx_j}, \ \ \ \ 
c_{j\downarrow}\to c_{j\downarrow} e^{-{i\over 2}Gx_j},
\end{equation}
upon which the Hamiltonian becomes translationally invariant (with a unit cell consisting of $L_y$ sites transverse to the adatom chain). The transformation in Eq.\ (\ref{hspi3}) has the effect of aligning the local spin quantization axis with the direction of $\bS_j$. The Hamiltonian becomes 
\begin{eqnarray}\label{hspi4}
{\cal H}_0&=&-\sum_{ij\sigma}t_{ij} e^{-{i\over 2}\sigma G(x_i-x_j)}c^\dagger_{i\sigma} c_{j\sigma} - \mu\sum_{i\sigma}  c^\dagger_{i\sigma} c_{i\sigma}\nonumber \\
 &&+JS\sum_{i\in I}  (c^\dagger_{i\sigma}\sigma^x_{\sigma\sigma'} c_{i\sigma'})
\end{eqnarray}
with the pairing term unchanged. In the spin reference frame rotating with the spiral the effective hopping term $\tilde{t}^\sigma_{ij}=t_{ij} e^{-{i\over 2}\sigma G(x_i-x_j)}$ is spin dependent and can be interpreted as containing an effective SOC. The last term in Eq.\ (\ref{hspi4}) represents a uniform Zeemann field of strength $JS$ pointed along the $\sigma^x$ direction on the adatom sites. 

In the 1D limit $L_y=1$ the transformed Hamiltonian represents a lattice version of the quantum wire model with SOC and superconducting order \cite{sau2,oreg1} that forms the theoretical basis for the existing experiments reporting Majorana zero modes \cite{mourik1,das1,deng1,rohkinson1,finck1,churchill1,Ramon}. To see the relation more clearly one can pass to the Fourier representation; the full Hamiltonian then becomes
\begin{eqnarray}\label{hspi6}
{\cal H}&=&\sum_q\bigl[\xi(q)c^\dagger_{q\sigma}c_{q\sigma}
+b(q)c^\dagger_{q\sigma}\sigma^z_{\sigma\sigma'}c_{q\sigma'} \\
&+& JSc^\dagger_{q\sigma}\sigma^x_{\sigma\sigma'}c_{q\sigma'}
+(\Delta c^\dagger_{q\uparrow} c^\dagger_{-q\downarrow}+{\rm h.c.})\bigr].\nonumber
\end{eqnarray}
In the above we have assumed a uniform SC order parameter along the chain, $\Delta_j=\Delta$ and defined quantities
\begin{eqnarray}\label{hspi6a}
\xi(q)&=&{1\over2}[\epsilon_0(q-G/2)+\epsilon_0(q+G/2)]-\mu, \\
b(q)&=&{1\over2}[\epsilon_0(q-G/2)-\epsilon_0(q+G/2)]
\end{eqnarray}
with $\epsilon_0(q)=-\sum_jt_{0j}e^{iqx_j}$.  The latter should be thought of as the normal-state dispersion in the absence of the exchange coupling and becomes simply $\epsilon_0(q)=-2t\cos{q}$ in the case of nearest neighbor hopping. When $J$ is non-zero the normal state-dispersion of ${\cal H}$ becomes
\begin{equation}\label{hspi7}
\epsilon(q)=\xi(q)\pm\sqrt{b(q)^2+J^2S^2},
\end{equation}
and the exchange coupling is seen to open a gap $2JS$ at $q=0, \pi$.
Assuming for simplicity that $t_{ij}=t$ for nearest neighbor sites and is zero otherwise the dispersion is plotted in Fig.\ \ref{fig2}c. 

An important feature to notice is that when the chemical potential $\mu$ lies inside the gap then there exists a single non-degenerate Fermi point in the right half of the Brillouin zone. According to the Kitaev criterion \cite{kitaev1} one expects the system to become a 1D topological superconductor upon the inclusion of the superconducting order. Combining Eq.\ (\ref{hspi7}) with (\ref{hspi6a}) we find that the chemical potential must satisfy
\begin{equation}\label{hspi8}
|\mu\pm\epsilon_0(G)|<JS
\end{equation}
for the system to be in the topological phase (assuming $\Delta$ to be small). If $G$ is considered a fixed parameter then the chemical potential must be adjusted rather accurately (e.g.\ by external gating) to lie in the magnetic gap. In practice, it is not clear how one would do this for adatoms on the surface of a superconductor. However, as pointed out in Refs.\ \onlinecite{simon1,loss1,vazifeh1}, $G$ should be viewed as a free parameter which adjusts itself so as to minimize the system free energy (or its ground state energy at $T=0$). It turns out that in 1D the equilibrium value of $G$ is given by 
\begin{equation}\label{hspi9}
\pm\epsilon_0(G)\approx\mu,
\end{equation}
implying that the condition (\ref{hspi8}) is always satisfied. Remarkably, in 1D the spiral pitch $G$ assumes that value which for a given chemical potential $\mu$ produces the topological phase \cite{simon1,loss1,vazifeh1}. This happens for all values of the chemical potential for which a solution of Eq.\ (\ref{hspi9}) for $G$ exists; the solution will not exist e.g.\ when $\mu$ lies outside of the band in which case the system is a band insulator.

As argued in Refs.\ \onlinecite{simon1,loss1,vazifeh1}, the reason behind this interesting behavior has to do with the form of the static spin susceptibility $\chi(q)$ of electrons in one dimension. Alternately, one can make a simple energy based argument appealing to the dispersion displayed in Fig.\ \ref{fig2}b,c. When the magnetic  gap opens up at $q=0$ the kinetic energy will be minimized when the occupied levels are pushed down in energy and empty levels are lifted up. A moment's reflection reveals that this happens precisely when for a given $G$ the chemical potential $\mu$ lies at the intersection of the $J=0$ bands marked as dashed lines in Fig.\ \ref{fig2}c. But this is exactly the condition indicated in Eq.\ (\ref{hspi9}). One can turn this around and see that for a given $\mu$ the spiral pitch $G$ minimizing the ground state energy will be the one satisfying the same Eq.\ (\ref{hspi9}).    

In this study we are interested in the fate of the self-organization phenomenon described above in the limit $L_y\gg 1$, corresponding to a 1D magnetic adatom chain positioned on a 2D superconducting surface. Since the translation invariance in the $y$ direction is now broken by the presence of adatoms it is no longer possible to find simple analytic forms for the quasiparticle dispersion in this case and we will rely primarily on numerical simulations. Also, as we shall see, the interplay between the magnetic ordering close to the chain and the bulk superconductivity away from it  will play an important role in the 2D problem and it is of key importance to treat the superconducting order parameter selfconsistently. The picture that emerges from these simulations is that of a 1D wire with predominantly magnetic order and a small SC gap embedded in a bulk 2D superconductor.

\section{Numerical results}

In this Section we present the results of our numerical simulations of the model defined by Hamiltonian Eq.\ (\ref{hspi1b}) in various 2D geometries and parameter regimes. We explore the system phase diagram, stability of the magnetic spiral with respect to disorder as well as the robustness of Majorana zero modes. 

\subsection{Periodic boundary conditions}

We start by considering Hamiltonian Eq.\ (\ref{hspi1b}) with the magnetic moments arranged in a spiral Eq.\ (\ref{hspi2}) on a $L_x\times L_y$ lattice with periodic boundary conditions along both $x$ and $y$. After the gauge transformation indicated in Eq.\ (\ref{hspi3}) we are thus led to the Hamiltonian   Eq.\ (\ref{hspi4}). The latter is translation invariant along the $x$ direction and it is convenient to perform a partial Fourier transformation 
\begin{equation}\label{num1}
c_{j\sigma}=\sum_k e^{ikx_j}c_{y_j\sigma}(k),
\end{equation}
where index $y_j$ now labels sites along the direction perpendicular to the magnetic chain. In the following we shall drop the subscript $j$ and label the the sites simply by $y$. In this representation the full BdG Hamiltonian becomes block diagonal in $k$ and can be written as 
\begin{equation}\label{num2}
\cH=\sum_k\sum_{y,y'}\Psi^\dagger_y(k)H_{yy'}(k)\Psi_{y'}(k) ,
\end{equation}
where $\Psi_y(k)=[c_{y\uparrow}(k),c_{y\downarrow}(k),c^\dagger_{y\downarrow}(-k),-c^\dagger_{y\uparrow}(-k)]^T$ is the Nambu spinor and each $H_{yy'}(k)$ is a $4\times 4$ matrix in the combined spin and Nambu spaces. The latter has the following structure
\begin{equation}\label{num3}
H_{yy'}(k)=
\begin{pmatrix}
h_{yy'}(k) & \delta_{yy'}\Delta_y \\
\delta_{yy'}\Delta_y^* & -\sigma^yh_{yy'}^*(-k)\sigma^y
\end{pmatrix},
\end{equation}
where $h_{yy'}(k)$ is the $2\times 2$ matrix (in the spin space) representing the normal state Hamiltonian (\ref{hspi4}) and $\Delta_y$ is the order parameter proportional to the unit matrix in the spin space.

\begin{figure}[b]
\includegraphics[width=8.0cm]{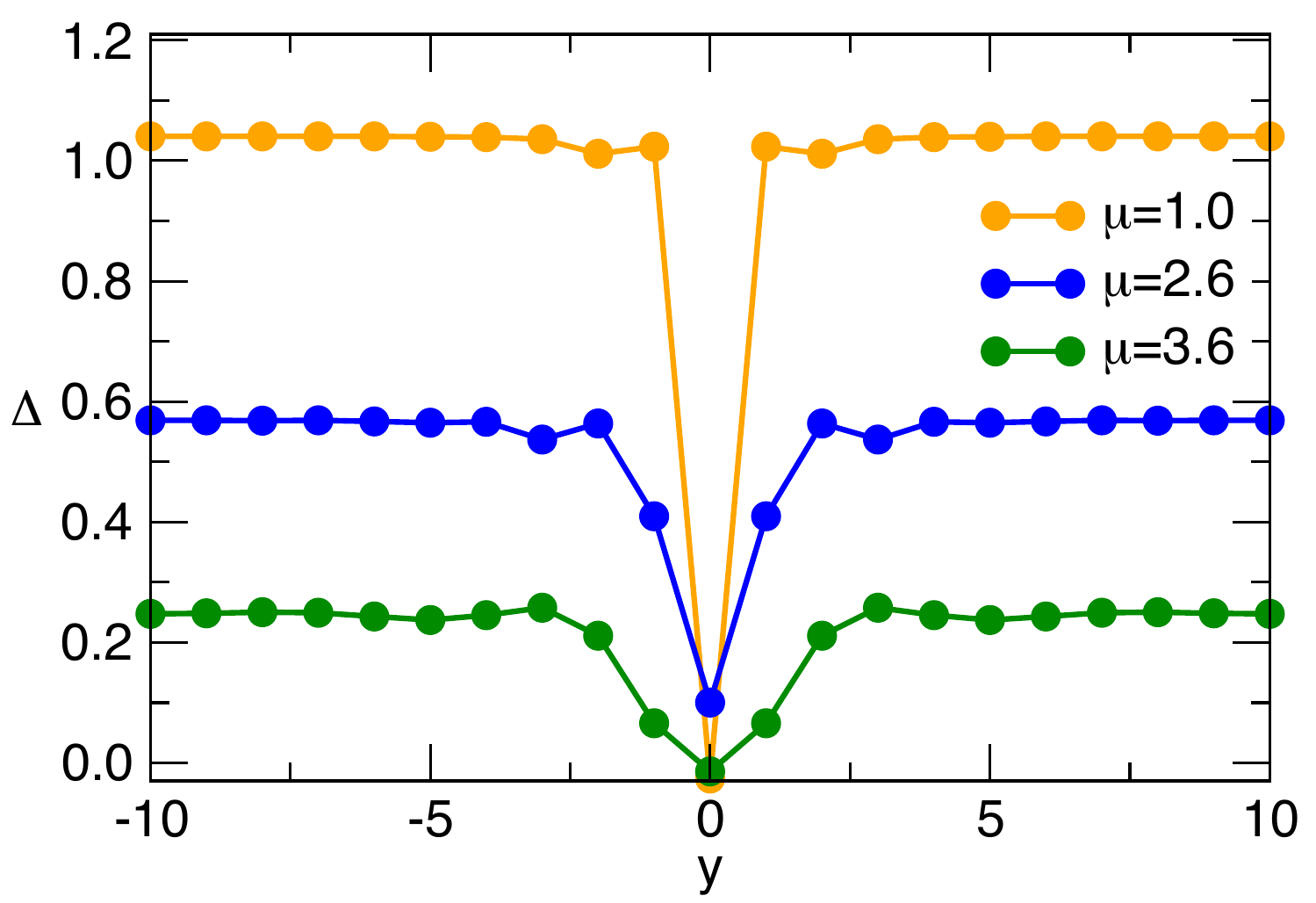}
\caption{Superconducting order parameter $\Delta_y$ obtained through the selfconsistent solution of Eq.\ (\ref{num4}) for a $100\times 20$ system with $V=3.6$, $JS=2$ and values of $\mu$ indicated in the legend. The spiral pitch used here corresponds to the global minimum of the energy ${\cal E}_g(G)$ defined in Eq.\ (\ref{num5}).  } \label{fig3}
\end{figure}
\begin{figure*}[t]
\includegraphics[width=16.7cm]{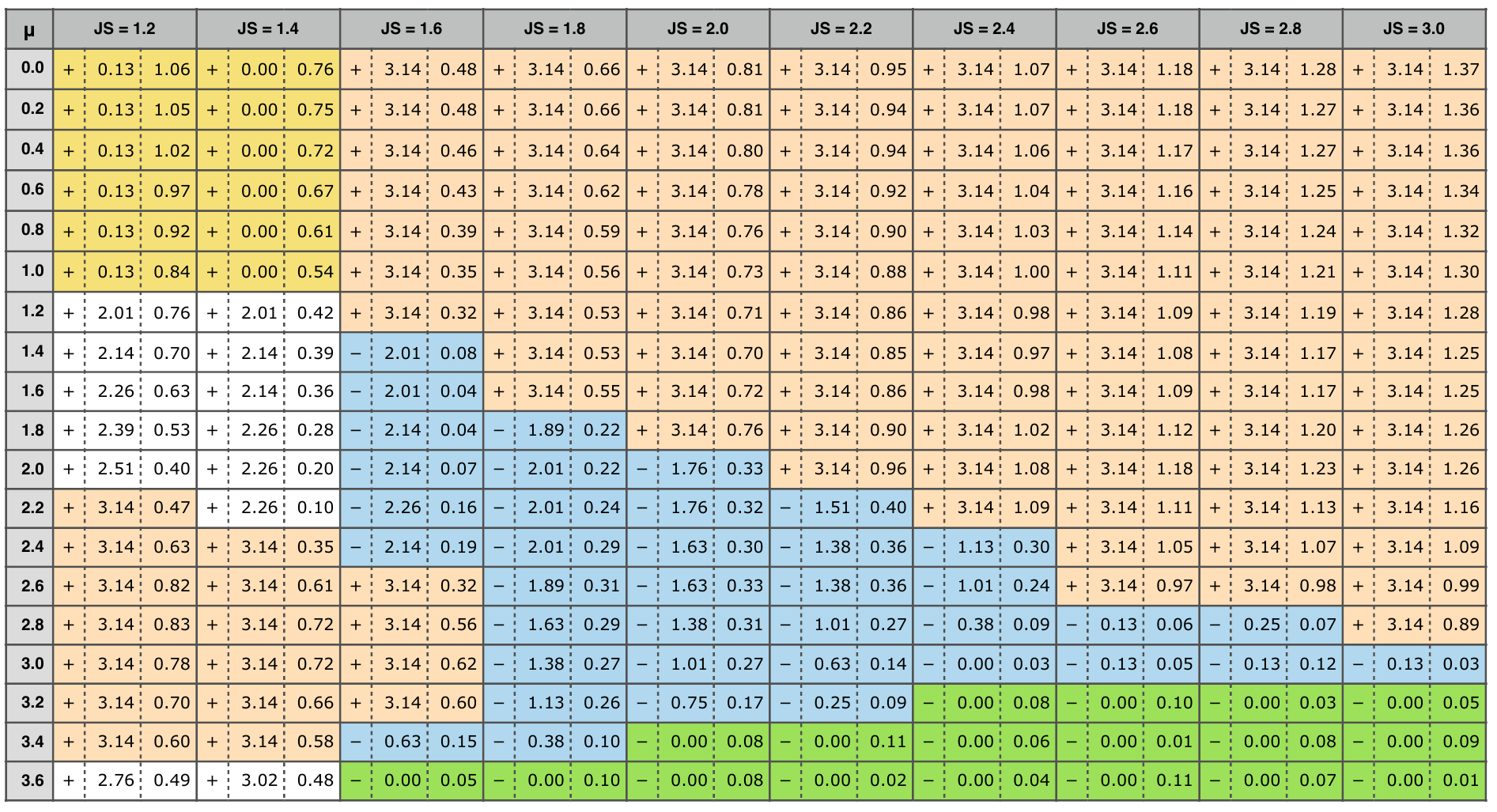}
\caption{ The phase diagram of the system with periodic boundary conditions described by Hamiltonian (\ref{num2}) for pairing strength $V=3.6$ and system size $L_x=100$, $L_y=20$. For each value of the chemical potential $\mu$ and the exchange coupling $JS$ we list three quantities obtained through the selfconsistent numerical procedure described in the text: the Majorana number ${\cal M}=\pm$, the spiral pitch $G_*$ and the quasiparticle excitation gap $\Delta_{\rm qp}$. The phases are as follows: gapped antiferromagnet (orange), gapped ferromagnet (yellow), topological spiral phase (blue), trivial spiral phase (white) and gapless ferromagnet (green).        } \label{fig4}
\end{figure*}
The problem at hand is solved by assuming an initial profile of the order parameter $\Delta_y$ and finding the eigenvectors and eigenvalues of $H_{yy'}(k)$,  
\begin{equation}\label{num4}
\sum_{y'}H_{yy'}(k)\Phi_{y'}(k) =E(k)\Phi_{y}(k),
\end{equation}
by exact numerical diagonalization. For each $k$ the size of the matrix to be diagonalized is $4L_y$. Given the set of $\Phi_{y}(k)$ and $E(k)$ one can then use the gap equation (\ref{hspi1c}) to compute the new order parameter profile $\Delta_y$ and iterate this procedure to selfconsistency.  Typical profiles of the superconducting order parameter $\Delta_y$ obtained by this procedure are displayed in Fig.\ \ref{fig3}. We observe a strong suppression of $\Delta_y$ along the magnetic chain. We also note that the strongest suppression occurs for the spiral pitch $G$ that minimizes the system energy, as discussed in more detail below. The important low energy degrees of freedom can be thought to live inside this potential well.

We perform this selfconsistent iteration for a fixed set of parameters $V$, $\mu$ and $JS$, expressed in units of the nearest neighbor hopping amplitude $t$ (which we set to 1) and cycle through all values of the spiral pitch $G$ consistent with the periodic boundary conditions. For each such $G$ we denote the corresponding manybody ground state by $|\Psi_G\rangle$ and evaluate the ground state energy 
\begin{equation}\label{num5}
{\cal E}_g(G)=\langle\Psi_G|\cH|\Psi_G\rangle.
\end{equation}
In the ground state and at suitably low temperatures $T$ we expect the magnetic moments to form a spiral at wavevector $G_*$ that minimizes ${\cal E}_g(G)$.

Written in the representation (\ref{num2}) our Hamiltonian can be viewed as describing a 1D system (along $x$) with $2L_y$ transverse bands. To determine whether or not the electrons form a topological phase we use Kitaev's criterion \cite{kitaev1} and compute the Majorana number given by
\begin{equation}\label{num6}
{\cal M}=\sgn[{\rm Pf}(\tilde{H}(0)){\rm Pf}(\tilde{H}(\pi))].
\end{equation}
Here Pf indicates the Pfaffian and $\tilde{H}(k)$ denotes the Hamiltonian matrix $H_{yy'}(k)$ written in the Majorana representation, i.e.\ as a purely imaginary, antisymmetric hermitian matrix. The structure of $H_{yy'}(k)$ displayed in Eq.\ (\ref{num3}) guarantees that a unitary transformation always exists that brings it to this form. ${\cal M}=-1$ indicates the topological phase in which unpaired Majorana zero modes are bound to the ends of the chain with open boundary conditions while  ${\cal M}=+1$ indicates the trivial phase \cite{kitaev1}.

Results of our numerical calculations for the system with periodic boundary conditions are summarized in Fig.\ \ref{fig4}. This should be viewed as a phase diagram in the space of the exchange coupling $J$ and the chemical potential $\mu$ for a fixed value of the pairing interaction strength $V=3.6$. Fig.\ \ref{fig4} shows one quadrant of the $J-\mu$ space but the remaining three quadrants can be obtained by simply reversing the signs of $J$ and $\mu$.
The phase diagram shares a number of features with the 1D system discussed previously \cite{vazifeh1}. For a given size of the SC gap the exchange coupling strength $JS$ must exceed a certain critical value to produce the spiral topological phase (rendered in blue). Close to the half filling $(\mu=0)$ the system tends to be antiferromagnetic while for the chemical potential close to the bottom of the band a gapless ferromagnetic phase prevails. Inside the topological phase the equilibrium spiral pitch $G_*$ evolves continuously as a function of $\mu$ and exhibits the self-organization property noted in connection with 1D systems \cite{simon1,loss1,vazifeh1}. There are also significant differences; while in 1D the topological phase extends to an arbitrarily large values of $JS$ here it occurs only over a limited range of exchange couplings $JS\simeq 1.6-3.0$. For larger values we find a direct transition from an AF insulator to an FM metal. For smaller values $JS<1.6$ we find an insulating FM state near the half filling followed by a topologically trivial spiral phase and an AF phase. In this regime the magnetic pair breaking effect is not sufficiently strong to significantly suppress $\Delta$ along the chain and create the effective 1D wire geometry. Consequently, the self-organization property fails to produce the topological phase.

\begin{figure}[t]
\includegraphics[width=8.7cm]{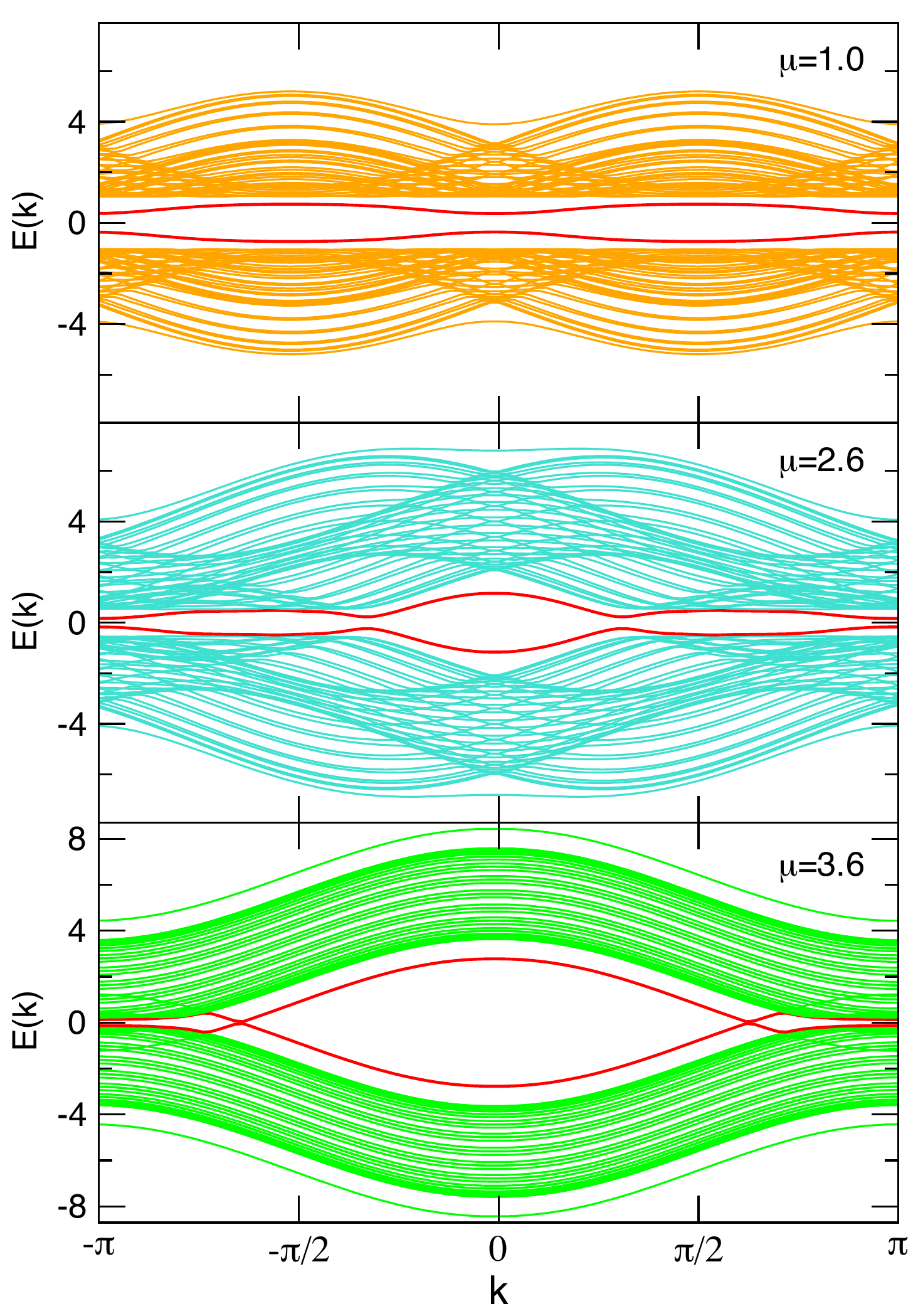}
\caption{Band structures $E(k)$ of the system obtained from Eq.\ (\ref{num4}) for a $100\times 20$ system with $V=3.6$, $JS=2$ and the spiral pitch $G_*$. The values of $\mu$ indicated in the legend correspond to the AF insulator (top), topological spiral phase (middle) and FM metal (bottom). The in-gap Shiba states are shown in red.  } \label{fig5}
\end{figure}

Several remarks are also in order. Most of the phase transitions indicated in  Fig.\ \ref{fig4} are first order with the spiral pitch undergoing a discontinuity. The exceptions are transitions from the topological phase to the trivial spiral phase (white) and gapless FM phase (green) which appear to be continuous within our resolution. The FM phase shows ${\cal M}=-1$ but a systematic investigation of larger system sizes indicates that it is a gapless metallic phase so the nontrivial topological invariant is not physically significant. In general excitation gaps $\Delta_{\rm qp}$ listed in Fig.\ \ref{fig4} that are smaller than $\sim 0.10$ are difficult to distinguish from the finite-size gaps inherent in our calculation. For this reason we have performed calculations with $L_x$ up to 1000 for some of the parameter values where the nature of the phase was in doubt. The assignment of phases in Fig.\ \ref{fig4} takes into account this additional work and we consider it reliable even when small values of $\Delta_{\rm qp}$ are displayed.  Band structures $E(k)$ typical of various phases are shown in Fig.\ (\ref{fig5}). In all cases we see a distinctive band of Shiba states associated with the magnetic moments with energies inside the bulk gap. The wavefunctions of these in-gap states are essentially one-dimensional with the largest amplitude along the chain and decay exponentially away from it. It is these Shiba states that can be described by the effective 1D model \cite{simon1,loss1,vazifeh1}  and give rise to Majorana zero modes in the geometry with open boundary conditions. 

\begin{figure*}[t]
\includegraphics[width=16.7cm]{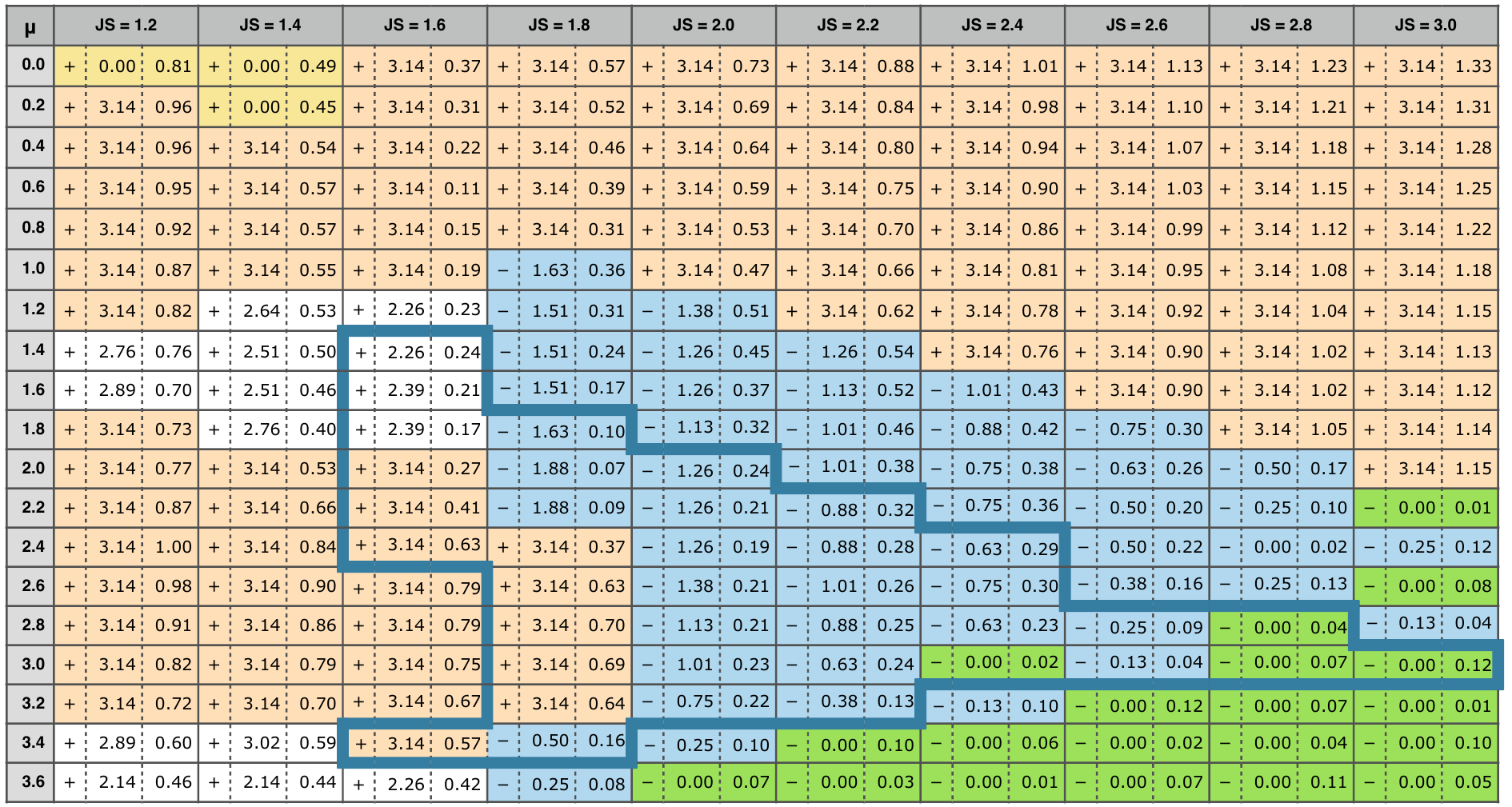}
\caption{ The phase diagram of the system with periodic boundary conditions. All parameters are the same as in Fig.\ \ref{fig4} except there is an additional scalar potential $\delta\mu=0.5$ on the magnetic sites as described by Eq.\ (\ref{num7}). The blue line outlines the topological phase from Fig.\ \ref{fig4} for comparison.} \label{fig6}
\end{figure*}

The phase diagram displayed in Fig.\ \ref{fig4} shows that in the 2D system under consideration the spiral topological phase discussed in the context of the 1D model \cite{simon1,loss1,vazifeh1} survives even though its extent is somewhat reduced compared to the 1D case. This is perhaps to be expected given the additional constraints imposed on the 2D system discussed in Ref.\ \onlinecite{bela1}. A question that one may ask now is how robust is the topological phase in Fig.\ \ref{fig4} with respect to changes of various system parameters. One could, for instance, consider more complicated electron band structures in the substrate superconductor and additional effects of the adatoms beyond the exchange coupling. Since the physics driving the spiral formation has to do with the low energy degrees of freedom (those below the bulk SC gap $\Delta$) we do not expect the substrate band structure to significantly affect the phase diagram. In the remainder of this subsection we thus explore the effect on the topological phase of the local change of the scalar potential imparted on the electrons by adatoms. We model this by a term 
\begin{equation}\label{num7}
\delta{\cal H}=-\delta\mu\sum_{i\in I}n_{i\sigma}
\end{equation}
that we add to the Hamiltonian (\ref{hspi1b}). We have explored the phase diagram for various values of $\delta\mu\in (-0.5,0.5)$ and found that although the shape, size and position of the topological phase are affected to some extent, the general topology of the phase diagram remains unchanged. Positive values of $\delta\mu$ tend to enlarge the topological phase while negative values reduce its extent. This can be understood by noting that positive $\delta\mu$ has the effect of depleting electrons from the chain and thus locally suppressing $\Delta(\br)$. This in turn produces a better defined  1D structure which, as we discussed, tends to support the topological phase. An example of the phase diagram modified by $\delta{\cal H}$ is given in Fig.\ \ref{fig6}.
%

\subsection{Open boundary conditions}

Simulations with open boundary conditions along $x$ are numerically more costly (one must diagonalize a matrix of the size $4L_xL_y$) but they allow one to directly investigate the Majorana zero modes as well as to address the effects of disorder. As an added bonus we can explicitly validate the assumption that the magnetic spiral represents the true ground state of the system. 

The simulations are performed directly in real space using Hamiltonian (\ref{hspi1b}) together with the gap equation (\ref{hspi1c}).
We consider disorder in on-site potential $\mu$ and in the exchange coupling $J$. To this end  in Eq.\  (\ref{hspi1}) we replace
\begin{equation}\label{num8a}
\mu\to\mu+\delta\mu_i,\ \ \ J\to J +\delta J_i
\end{equation}
where $\delta\mu_i$ and $\delta J_i$ are independent random variables taken on each site from the interval $(-w_\mu,w_\mu)$ and $(-w_J,w_J)$, respectively, with a constant probability. In addition we do not assume a spiral phase but treat each magnetic moment as an independent classical fluctuating degree of freedom whose dynamics is controlled by its coupling to the substrate electrons. To keep the simulations manageable we confine the moments to rotate in a single plane and allow only discrete orientations 
\begin{equation}\label{num8}
\bS_i=S\left(\cos{\theta_i},\sin{\theta_i},0\right),  \ \ \
\theta_i={2\pi m_i\over M}
\end{equation}
with $m_i=0,1 \dots M-1$ and $M$ a large integer. A set of integers $\{m_i\}$ thus specifies the magnetic state of the chain. 

We find the ground state by the method of simulated annealing. Starting from a random moment configuration $\{m_i\}$ we find the SC order parameter as described in the previous subsection and compute the system energy $\cE_g$. We then choose a magnetic moment $i$ at random and change $m_i\to m_i\pm 1$.  Ground state energy $\cE_g'$ for this new configuration is then computed, as above. The update is accepted or rejected according to the standard Metropolis algorithm:   
\begin{eqnarray}\label{num9}\nonumber
{\rm if}  &&\cE_g'< \cE_g \ \ \ {\rm accept}\\
{\rm if}  &&\cE_g'> \cE_g \ \ \ {\rm accept\ with \ probability} \ \ e^{-(\cE_g'-\cE_g)/T} \nonumber
\end{eqnarray}
where $T$ is a (fictitious) temperature parameter which we slowly lower during the annealing. This way we find the ground state of the system (or  one of the low-lying metastable states) without any bias towards a particular magnetic state, aside from assuming a coplanar ordering of moments. 
\begin{figure*}
\includegraphics[width=15.7cm]{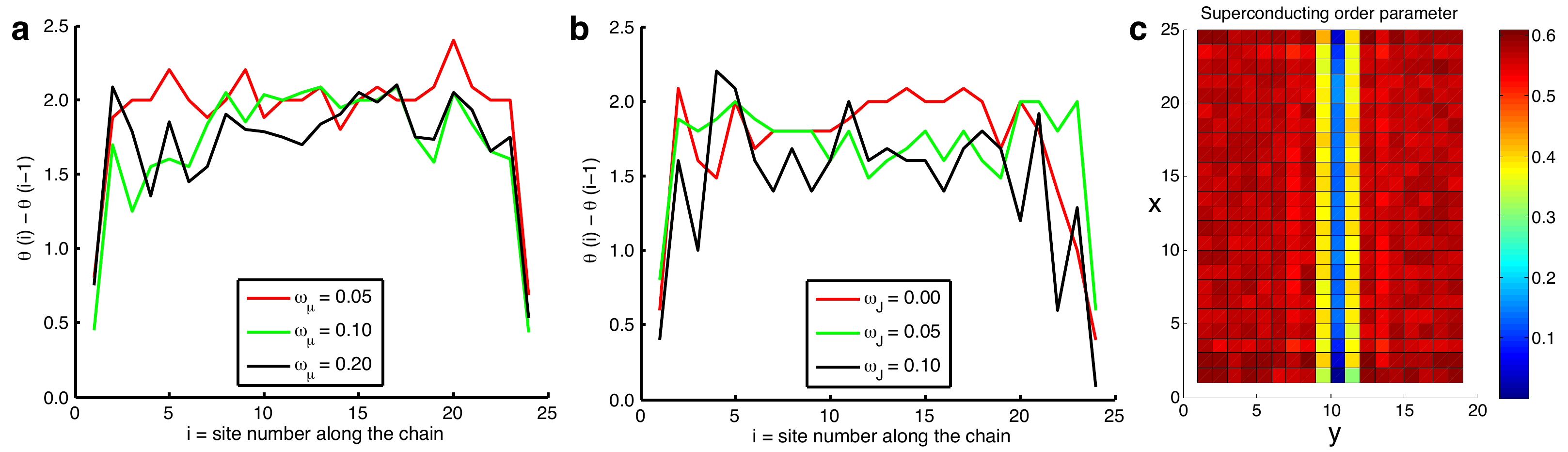}
\caption{Magnetic spiral for various levels of disorder in a) the on-site potential $\mu$ and b) the exchange coupling $J$. Difference between  spiral angles $\theta_i$ on neighboring sites are plotted for system size $L_x=25$, $L_y=19$ and parameters $V=3.6$, $JS=2$ and $\mu=2.6$. We used $M=31$ for $\theta_i$ increment. c) Density plot of the SC order parameter for the same parameters as above and $w_\mu=0.05$. } \label{fig7}
\end{figure*}

Fig.\ \ref{fig7} shows some representative results of our simulated annealing calculations, performed for parameters expected to yield the topological phase. The magnetic state is represented by plotting differences $\Delta\theta_i=\theta_i-\theta_{i-1}$ between the moment angles on neighboring sites. For a perfect spiral these would be constant and independent of the site index $i$. In panels (a) and (b) we observe that the system ground state is close to a perfect spiral with fluctuations in  $\Delta\theta_i$ increasing somewhat as we ramp up the disorder. We note that our annealing procedure yields some fluctuations away from the perfect spiral even for the clean case. We attribute these to the fact that we allow only discrete values of $\theta_i$ in our simulations which can only approximate the perfect spiral unless its pitch is commensurate with $2\pi/M$. Nevertheless these results indicate that a magnetic state close to the perfect spiral is obtainend in an unbiased calculation, even in the presence of moderate amount of disorder in the on-site potential $\mu$ and the exchange coupling $J$. The spiral is somewhat more sensitive to disorder in $J$ which is to be expected. We also note that the spiral tends to be altered near the ends of the chain. This too is to be expected because various symmetries are broken near the edges.

 For stronger levels of disorder it becomes more difficult to equilibrate the system and our simulations become less reliable. We often find domain walls in the spiral when the disorder is strong. These are manifestations of the fact that in the clean system spiral order at $G$ is degenerate with order at $-G$. Disorder breaks this degeneracy locally and causes, presumably, proliferation of domain walls which eventually destroy the spiral order. We note that according to Ref.\ \onlinecite{ojanen1} such domain walls harbor protected pairs of Majorana zero modes and we have indeed observed these in our simulations.  

\begin{figure}[b]
\includegraphics[width=8.7cm]{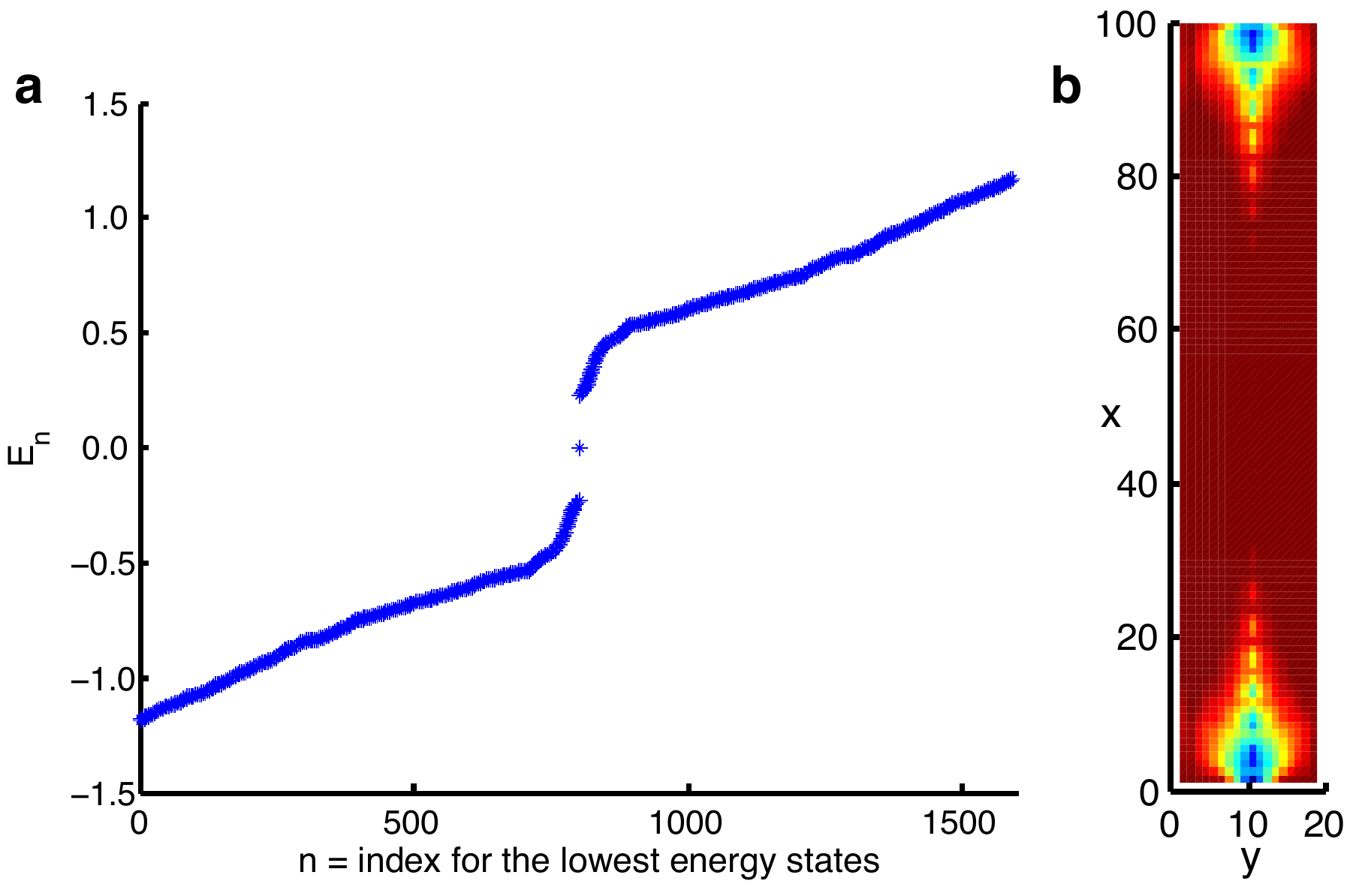}
\caption{a) Energy spectrum of the $100\times 19$ system with $w_\mu=0.05$ and disorder in spiral order as described in the text. b) Density plot of the lowest positive energy wavefunction amplitude representing the symmetric combination of the Majorana endmodes. Note that the false color is on logarithmic scale to achieve better visual contrast. The system parameters are as in Fig. \ref{fig7}.} \label{fig8}
\end{figure}
An important question that we wish to address is the stability of the Majorana endmodes with respect to disorder. Since for moderate amounts of disorder the spiral order remains globally stable we expect the Majoranas to also remain robust. For the relatively small system sizes for which we can reliably perform the simulated annealing procedure, however, it is difficult to directly study this question because of the significant overlap of the Majorana wavefunction leading to the zero mode energy splitting. (We note however that the splitting occurs already in the clean systems of this size and is not significantly altered by disorder). For this reason we consider a larger system ($L_x=100$, $L_y=19$) with disorder in $\mu$. Instead of finding the true equilibrium moment configuration $\{m_i\}$, which would be impractical for a system of this size, we impose a moment configuration with fluctuations away from the perfect spiral that are {\em statistically} the same as we found in a smaller system with this level of disorder. The results are displayed in Fig.\
\ref{fig8} and indicate a pair of zero energy modes well separated from the rest of the states. The wavefunctions corresponding to these zero modes are localized near the ends of the magnetic chain and correspond to the Majorana endmodes expected to be present in the topological phase.

\section{Summary and conclusions}

Adatoms with uncompensated magnetic moments deposited on the surface of an ordinary $s$-wave superconductor act as strong pairbreakers because of their exchange coupling to the electron spins. This causes local suppression of the SC order parameter and the emergence of Shiba states inside the gap with wavefunctions localized in the vicinity of the moments. When such adatoms are arranged to form a line then the Shiba states hybridize and create a 1D band of in-gap states. At low energies, this band can be thought of as representing a 1D wire in which magnetic and SC orders compete. Our selfconsistent calculations based on the 2D lattice Hamiltonian (\ref{hspi1a}) find that the magnetic moments tend to arrange in a coplanar spiral with an equilibrium pitch $G_*$ that depends on the system parameters. For a range of parameters that can be expected to occur in realistic systems (e.g.\ Gd atoms on a Pb surface \cite{yazdani00}) we find that $G_*$ adjusts itself such that the resulting wire forms a 1D topological superconductor with Majorana zero modes bound to its ends. The self-organization property found previously in purely 1D models \cite{simon1,loss1,vazifeh1} is thus recovered over a portion of the phase diagram in this more realistic 2D model. We emphasize that the topological phase occurs here in a fully SU(2) symmetric model -- SOC is not required to bring it about (although a small amount of SOC may be needed to stabilize the spiral against thermal fluctuations \cite{vazifeh1}, not considered in this work). 

An even more realistic model would include a full 3D description of the SC substrate. A fully selfconsistent calculation along the lines presented above  is currently out of our reach due to a significant numerical cost incurred in modeling a sufficiently large 3D system. We expect, however, that the same mechanism that we uncovered in the 2D model will operate in 3D and the self-organization property should carry over largely intact.   

\acknowledgements
The authors are indebted to J. Alicea, B. Bauer, L.I. Glazman, G. Refael, M.M. Vazifeh and A. Yazdani for illuminating discussions and correspondence. The work presented here was supported in part by NSERC and CIfAR. M.F.\ thanks The Institute for Quantum Information and Matter at the California Institute of Technology for hospitality while this work was finalized.


\end{document}